# Mechanics of diffusion-mediated budding and implications for virus replication and infection

Mattia Bacca

*Mechanical Engineering Department, School of Biomedical Engineering, Institute of Applied Mathematics*
*University of British Columbia, Vancouver BC V6T1Z4, Canada*

*E-mail address:* mbacca@mech.ubc.ca

**Abstract**

Budding allows virus replication and macromolecular secretion in cells through the formation of a membrane protrusion (bud) that evolves into an envelope. The largest energetic barrier to bud formation is membrane deflection and is trespassed primarily thanks to nucleocapsid-membrane adhesion. Transmembrane proteins (TPs), which later form the virus ligands, are the main promotors of adhesion and can accommodate membrane bending thanks to an induced spontaneous curvature. Adhesive TPs must diffuse across the membrane from remote regions to gather on the bud surface, thus, diffusivity controls the kinetics. This paper proposes a simple model to describe diffusion-mediated budding unraveling important size limitations and size-dependent kinetics. The predicted optimal virion radius, giving the fastest budding, is validated against experiments for Coronavirus, HIV, Flu, and Hepatitis. Assuming exponential replication of virions and hereditary size, the model can predict the size distribution of a virus population. This is verified against experiments for SARS-CoV-2. All the above comparisons rely on the premise that budding poses the tightest size constraint. This is true in most cases, as demonstrated in this paper, where the proposed model is extended to describe virus infection via receptor- and clathrin-mediated endocytosis, and via membrane fusion.

## 1. Introduction

Enveloped viruses are ubiquitous in nature. They have covered a fundamental role in the evolution of the living kingdom and can have a tremendously disruptive impact on human health and economy, as observed in recent times. While subject to intense study from the biochemical point of view, the physical mechanisms involved in virus replication and infection have comparatively received rather limited attention [1-2]. Despite the controversy around the definition of a virus as a form of *life*, it is commonly observed that viruses replicate and *evolve* to optimize replication at the planetary scale [2]. Replication is preceded by the infection of a host cell [2], thus, infection must occur for a virus to replicate itself. This explains why infection has been the center of attention in recent times. However, as demonstrated in this paper, and in agreement with observations in the literature, replication is a much slower biophysical process than infection. Hence, it can cover a more significant role in the life cycle of a virus.

Enveloped viruses are characterized by a lipid membrane wrap (envelope) surrounding the nucleocapsid, which encapsulates genetic material [2] (RNA). The membrane is decorated with transmembrane proteins (TPs), of which, the most important ones are virus ligands (spike proteins). Virus ligands are macromolecular assemblies that protrude from the virion (virus particle) and anchor to the receptors of the host cell to prompt infection. The latter involves RNA delivery inside the cytosol of the infected cell (host). The virulent RNA can hijack the host's protein-duplication mechanism to replicate the components of the virus. The new components then assemble into new nucleocapsids inside the cytosol, while the new TPs are delivered to the host membrane. Virus replication concludes with budding, the process by which the nucleocapsid wraps around the host membrane, equipped with virus ligands, and is ultimately expelled from the mother cell into a new enveloped virion [2-3] (see Figure 1). Budding requires significant membrane deflection, and this constitutes the main energetic barrier to the process. Because TPs protrude from the cell membrane, pointing outward, they create a local spontaneous curvature [4-7], which facilitates wrapping. The energetic barrier becomes then a function of the radius of the capsid, relative to the spontaneous membrane curvature, with the latter defining the optimal capsid radius for fast replication. This also provides minimum and maximum radii ensuring the spontaneity of budding. To accommodate membrane bending, the TPs have to diffuse in the location of the bud. Their diffusion is promoted by their high affinity with the nucleocapsid (cargo); thus, TP diffusion controls the kinetics of the process [8]. [8] developed the first theoretical model to describe budding from an energetic analysis. The free energy of the bud includes the bending energy of the membrane and the binding energy with the nucleocapsid, mediated by spike proteins. The authors also speculated on the possibility of the spike proteins accommodating non-zero spontaneous curvature in the membrane, thereby lowering the energetic barrier for bud formation. They estimated the budding time as comprised between 10 and 20 minutes, in agreement with prior measurements. However, they did not explore the influence of virion size in the process. [9] continued this investigation by providing a steady-state model for budding incorporating the many-body interaction among multiple forming buds. Here, the free energy of the system includes membrane bending, spike adhesion, and the line energy of the bud rim. They provided the energetic landscape of the multi-bud system and discussed the implications of size and spike density in budding, but did not consider the spontaneous membrane curvature induced by the TPs, nor the kinetics of TP diffusion.

Both the above-discussed investigations provided important basic principles to better understand the physics of virus replication, however, none provided a quantitative investigation of the role of virion size on budding spontaneity and kinetics. This paper provides a simple model to describe the process, inspired by a simple model for diffusion-mediated endocytosis [3]. It provides virion size limitations via energetic analysis and size-dependent budding kinetics. The model ultimately finds the optimal virion size at which budding is fastest. This optimum is compared against experiments for several virus species, namely SARS-CoV-1 & 2 [10-12] (coronavirus), HIV-1 (HIV type 1) [13], Flu (Influenza) [14-15], and HCV (Hepatitis type C) [16-17]. Using an exponential replication model based on size hereditariness, the proposed budding model can predict the size polydispersity of a virus population. This prediction is then compared against experiments for SARS-CoV-2 (novel coronavirus) [11] as an example. The reader will find it easy to create such a comparison for all other virus species. The above comparisons are based on the hypothesis that budding is the most size-limiting process in the life cycle of a virus. This hypothesis is later validated by investigating the size limitations of virus infection; finding that, for the majority of biologically relevant cases, budding imposes the tightest size constraints on the virion. Virus infection is studied considering the two main infection mechanisms, namely, receptor- and clathrin-mediated endocytosis [3,18] and membrane fusion.

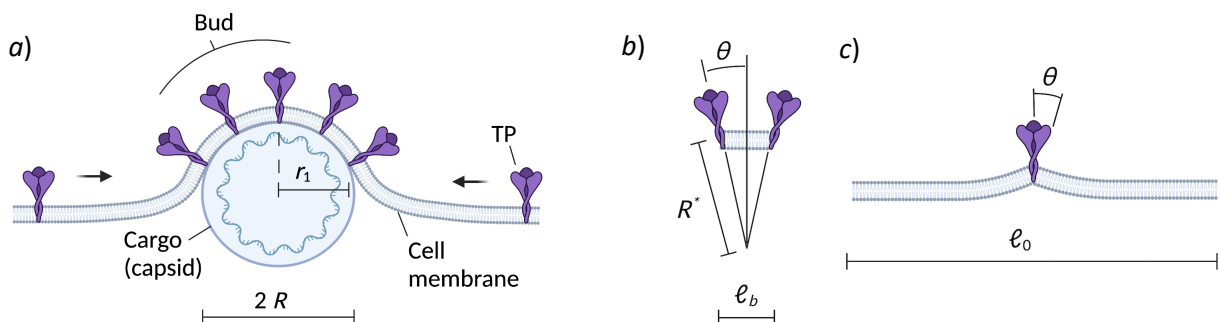

**Figure 1**: Schematics of the (virus) budding process: *a)* The cell membrane adheres with the cargo (virus capsid) through transmembrane protein (TP) interaction; *b)* TPs induce a spontaneous curvature $1/R^*$ to the budding membrane, which is function of the conical TP angle $\theta$ and TP spacing $\ell_b$ on the bud surface; *c)* The TP also induces local bending in the rest (flat) membrane of the cell due to the angle $\theta$, with $\ell_0$ the TP spacing in the resting membrane.

The model can predict size constraints and the polydispersity of a virus population from molecular scale properties such as TP adhesion, spontaneous membrane curvature, and TP availability. It can be also used in reverse, where virus size polydispersity can be used to predict molecular scale properties. This is particularly useful to extrapolate the value of parameters that are difficult to measure, such as TP-capsid adhesion.

## 2. Budding mechanics model

As sketched in Figure 1a, in the proposed model, the kinetics of the process is controlled by TP diffusion, as observed by [8]. The free energy of the system $\psi$, in its dimensionless form, is

$$\frac{\psi}{kT} = \left[-g_m - \rho_b e_T + \sigma \varepsilon_S + \rho_b \ln\left(\frac{\rho_b}{\rho_0}\right) + 2B\left(\frac{1}{R} - \frac{1}{R^*}\right)^2\right] S_b + 2\pi \int_{r_1}^{\infty} \rho \ln\left(\frac{\rho}{\rho_0}\right) r dr \quad (1)$$

Here $k$ is the Boltzmann's constant; $T$ is temperature; $g_m$ is the adhesion surface energy between the cargo surface (CS, the virus nucleocapsid) and the cell membrane; $e_T$ is the energy released by a TP joining the bud; $\sigma$ is the surface tension of the cell membrane, here considered constant for simplicity albeit dependent on membrane curvature [7]; $\varepsilon_S = \Delta S/S_b$ is the surface strain of the membrane, with $\Delta S$ the change in membrane surface and $S_b$ the bud surface; $\rho$ is the TP density in the membrane, with $\rho_0$ and $\rho_b \approx l_b^{-2}$ [18] that in the resting membrane and in the bud (*i.e.* the ligand density in the virion), respectively, and $l_b$ the TP (ligand) spacing in the bud (virus particle); $B$ is the bending modulus of the membrane; $R$ is the radius of curvature of the CS, *i.e.* that of the wrapping membrane, with $R^*$ the spontaneous radius of curvature of the membrane; $r_1$ is the distance between the axis of symmetry and the *bud rim* (the point of contact between the cargo and the membrane, see Figure 1a). Eq. (1) neglects the bending energy of the membrane outside the bud and the line energy of the bud rim. A full elastic solution would provide completeness [19], however, [4] showed that the cell membrane in that region assumes a zero-energy catenoid-like configuration. This is also proven by the calculations of [20], where the energy of the system is nearly-constant during wrapping. $\varepsilon_S > 0$ constitutes an additional energy penalty due to stretching of the cell membrane, *i.e.* the diffusion of amphiphiles to compensate for the surface area change during budding. $\varepsilon_S < 0$ provides an energetic driving force due to an excess of amphiphiles in the membrane, which can be removed by expelling a vesicle (membrane buckling).

The spontaneous radius of curvature $R^*$ is given by the conical angle $\theta$ created by the TPs located in the bud, as shown in Figure 1b, giving

$$\frac{1}{R^*} = 2\theta\sqrt{\rho_b} \quad (2)$$

The energy released by one TP joining the bud is

$$e_T = e_{T_a} + e_{Tm} \quad (3a)$$

with $e_{T_a}$ the adhesion energy between a TP and the CS, and

$$e_{Tm} = B 2\pi \theta^2 \frac{d_{TP}^2}{\ell_0^2 - d_{TP}^2} \quad (3b)$$

the bending energy associated with the local curvature created by the TP in the resting membrane, as depicted in Figure 1c [4]. This energy is released once the TP joins the bud, hence it provs a driving force. In Eq. (3b), $d_{TP}$ is the diameter of the TP and $\ell_0$ is the TP spacing in the resting membrane.

*Appendix A* provides the solution to the transient problem, where the bud rim moves following the kinetic law

$$r_1 = 2\alpha\sqrt{Dt} \quad (4)$$

with $\alpha$ a kinetic constant called 'speed factor' [3], $D$ the diffusivity of a TP in the membrane, and $t$ time. $D$ can be calculated using the bidimensional Stokes-Einstein relation [21], assuming the lipid membrane behaves like a fluid mosaic, giving $D = kT/1.69\pi\eta$. Here, $\eta = 10^{-5} N\,s/m$ is the viscosity of the membrane measured via rheology [22]. Also note that in the above relation, $D$ is independent of $d_{TP}$. The value of $D$, obtained via bidimensional Stokes-Einstein relation, is provided in Table 1 and is in agreement with the value adopted by [3].

Eq. (4) can be inverted to give the time required for budding completion (budding time), $t_b$, from the relation $S_b = \pi r_1^2 = 4\pi R^2$, thus

$$t_b = \frac{R^2}{D\alpha^2} \tag{5}$$

To calculate $t_b$, one has to compute $\alpha$. This is reported in *Appendix A* from the condition

$$e_b \geq F_{\tilde{\rho}}(\alpha) \tag{6a}$$

where $e_b$ is the *budding energy*, i.e. the driving force for budding, and

$$F_{\tilde{\rho}}(\alpha) = (1-\tilde{\rho})f(\alpha) - \ln\left[1 + \left(\frac{1}{\tilde{\rho}} - 1\right)f(\alpha)\right] \tag{6b}$$

is the kinetic function of the process, with $\tilde{\rho} = \rho_0/\rho_b$ the dimensionless equilibrium density of TPs in the resting membrane, and

$$f(\alpha) = \frac{\alpha^2 E(\alpha^2)}{\alpha^2 E(\alpha^2) - \exp(-\alpha^2)} \tag{6c}$$

with $E(\alpha^2) = \int_{\alpha^2}^{\infty} e^{-u} du/u$ the exponential integral. The budding energy in Eq. (6a) is given by

$$e_b = e_{b0} - 2\tilde{B}\left(1 - \frac{R^*}{R}\right)^2 \tag{7a}$$

where

$$\tilde{B} = \frac{B}{\rho_b R^{*2}} = 4B\theta^2 \tag{7b}$$

is the dimensionless bending rigidity of the membrane, and

$$e_{b0} = e_{Ta} + e_{Tr} + e_{Tm} + \tilde{g}_m - \tilde{\sigma}\varepsilon_S \tag{7c}$$

is the *budding input energy*, independent of the cargo size ($R$), with

$$e_{Tr} = 1 - \tilde{\rho} + \ln\tilde{\rho} \tag{7d}$$

the energetic cost of TP relocation, and $e_{Tm}$ the released bending energy given by Eq. (3b) and rewritten as

$$e_{Tm} = \tilde{B}\frac{\pi}{2}\frac{\xi\tilde{\rho}}{1-\xi\tilde{\rho}} \tag{7e}$$

with

$$\xi = \frac{\sqrt{3}}{2}d_{TP}^2\rho_b \tag{7f}$$

Eq. (7e-f) are derived by adopting a hexagonal distribution of TP in the resting membrane, for which $1/\rho_0 = \sqrt{3}\ell_0^2/2$.

From Eq. (6) and (7), it should be noted that $\alpha$ is proportional to $e_b$, so that a higher driving force $e_b$ can produce faster budding or smaller $t_b$. Conversely, $\alpha = 0$, i.e. $t_b \to \infty$, provides the *critical condition* for budding spontaneity. From Eq. (6a), we have that $F_{\tilde{\rho}}(0) = 0$ for any $\tilde{\rho}$, thus $e_b \geq 0$ gives the necessary condition for spontaneous budding. This condition, applied to Eq. (7a), provides the size constraints

$$R_{min} \leq R \leq R_{max} \tag{8a}$$

With

$$\frac{R_{max}}{R^*} = \frac{1}{1-\sqrt{e_{b0}/2\tilde{B}}} \tag{8b}$$

$$\frac{R_{min}}{R^*} = \frac{1}{1+\sqrt{e_{b0}/2\tilde{B}}} \tag{8c}$$

It should be noted that Eq. (8b) only applies for $e_{b0} < 2\tilde{B}$, whereas for higher input energy $R_{max} \to \infty$ and Eq. (8) only provides a constraint for minimum radius.

The inequality in Eq. (8a) ensures budding spontaneity. The next section reports the budding time as a function of $e_{b0}$, $\tilde{\rho}$, and $R$.

Table 1 reports the parameter values adopted in this investigation and their source, while the next section explores the size limitations and size-dependent kinetics of the problem.

## 3. Results & Discussion

*Budding in absence of TPs*

In the absence of TPs, the kinetics of the process is controlled by the relaxation time of the membrane. Here, the model is simply used to define the size constraints for budding spontaneity, neglecting the kinetics of budding. In this case, $\rho, \rho_b, \rho_0, 1/R^*$ all vanish to zero, thus, Eq. (6) rewrites as

$$g_m - \sigma\varepsilon_S \geq \frac{2B}{R^2} \tag{9}$$

| Parameter | Source | Value (range; adopted median) |
|---|---|---|
| $kT$ | [23] | $4 \cdot 10^{-21} J$ |
| $\rho_b$ | [3] | $3 \cdot 10^{-3} - 20 \cdot 10^{-3} \, nm^{-2}; 5 \cdot 10^{-3} \, nm^{-2}$ |
| $l_b$ | $\rho_b^{-1/2}$, [18] | $7 - 18 \, nm; 14 \, nm$ |
| $\sigma$ | [24] | $5 \cdot 10^{-3} \, nm^{-2}$ |
| $B$ | [3] | $10 - 25; 20$ |
| $\theta$ | [11] | $\theta \approx 8.2° = 0.14 \, rad$ |
| $R^*$ | Eq. (2) | $50 \, nm$ |
| $d_{TP}$ | [12] | $4 \, nm$ |
| $D$ | [3] | $10^4 \, nm^2/s$ |
| $\tilde{\rho} = \rho_0/\rho_b$ | [3,18] | $0.01 - 0.1$ |
| $\tilde{B}$ | Eq. (7b) | $1.64$ |
| $e_{b0}/2\tilde{B}$ | Figure 3 | $0.03 - 0.2; 0.05$ |
| $e_{b0}$ | | $0.1 - 0.66; 0.16$ |
| $e_{RL}$ | [3] | $10 - 25; 20$ |
| $B_c$ | [25] | $255 - 315; 300$ |
| $B_c/B$ | | $10.2 - 31.5; 15$ |
| $\rho_c$ | [25] | $1.25 \cdot 10^{-3} nm^{-2}$ |
| $\tilde{\rho}_c = \rho_c/\rho_b$ | | $0.25$ |
| $e_c$ | [25] | $5 - 30; 23$ |
| $R_c$ | [25] | $32.5 - 90 \, nm; 50 \, nm$ |

**Table 1**: Parameters adopted in the model, together with their source, value range and adopted median value.

From this, one can obtain the minimum (dimensionless) bud radius as

$$\frac{R_{min}}{\sqrt{2B/\sigma}} = \frac{1}{\sqrt{g_m/\sigma - \varepsilon_S}} \tag{10}$$

Figure 2 plots $R_{min}/\sqrt{2B/\sigma}$ versus the dimensionless adhesion energy $g_m/\sigma$ and surface strain $\varepsilon_S$, from Eq. (10). All plots show a horizontal asymptote for very strong adhesion, leading to a zero minimum radius. *I.e.*, in the case of very strong adhesion, compared to membrane tension, or $g_m \gg \sigma\varepsilon_S$, we have that $R_{min} \ll \sqrt{2B}$. The minimum radius also appears to be highly sensitive to the surface strain $\varepsilon_S$. When $\varepsilon_S > 0$, *e.g.* in the case of $\varepsilon_S = 1$, the energy of surface stretch, together with bending energy, has to be compensated by adhesion. In the case of $\varepsilon_S = -1$, the surface energy in excess is equivalent to the surface of the bud. In the absence of adhesion, the latter case is the only one that can favor budding since for $g_m = 0$, all other plots give $R_{min} \to \infty$ (*i.e.* no budding spontaneity for any size). The case of $g_m = 0$ and $\varepsilon_S = -1$ is also that in which the cargo can be absent, *i.e.* vesiculation, where $R_{min} = \sqrt{2B/\sigma}$. In this case, the excess surface energy is released in the form of bending energy, *i.e.* membrane buckling. Taking the values reported in Table 1 for $B$ and $\sigma$, the minimum vesicle radius computes to $R_{min} = 60 - 100\ nm$, in agreement with experimental observations [26].

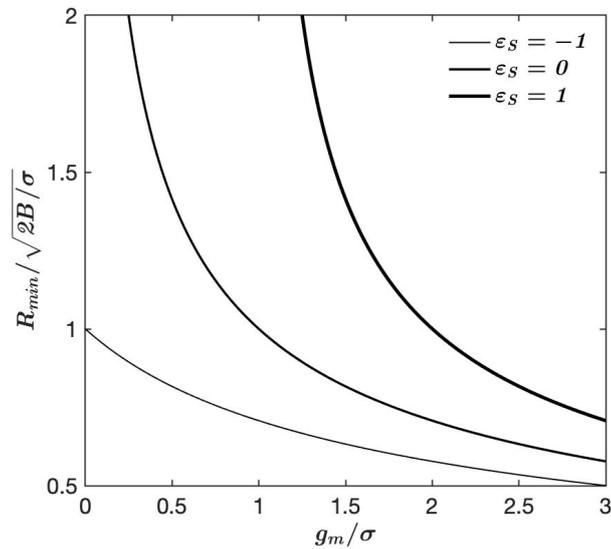

**Figure 2**: Budding in the absence of TPs. The plots report dimensionless minimum bud radius versus dimensionless adhesion energy at various surface strains, from Eq. (10).

*TP-mediated budding*

In the presence of TPs one can compute the minimum and maximum radii of curvature of the cargo allowing spontaneous budding (*i.e.* the size constraints of the virion), from Eq. (8). Figure 3a plots these radii as a function of the ratio $e_{b0}/2\tilde{B}$. $\tilde{B}$ is estimated from Eq. (7b) and the parameter values in Table 1, and its value is reported in the same table. To the author's knowledge, $e_{b0}$ has never been measured and is here estimated from the observed maximum/minimum radii. Figure 3b reports the predicted $e_{b0}/2\tilde{B}$ as a function of the ratio $R_{max}/R_{min}$, from Eq. (8), and from the maximum/minimum radii extracted from experimental observations of several virus species. These are SARS-CoV-1 & 2 (coronavirus) [10-12], HIV-1 (human immunodeficiency virus type 1) [13], Flu (influenza) [14-15], and HCV (hepatitis C virus) [16-17]. As evidenced in this figure, the proposed model predicts the range $e_{b0}/2\tilde{B} = 0.03 - 0.2$, while the majority of data points suggest a median value of 0.05. From Eq. (7c) one can deduce that the variability of input energy within the same virus species is likely to be attributed to the stochastic variation of TP concentration (variable $\tilde{\rho}$) and/or the stochastic presence of excess amphiphiles (variable $\varepsilon_S$).

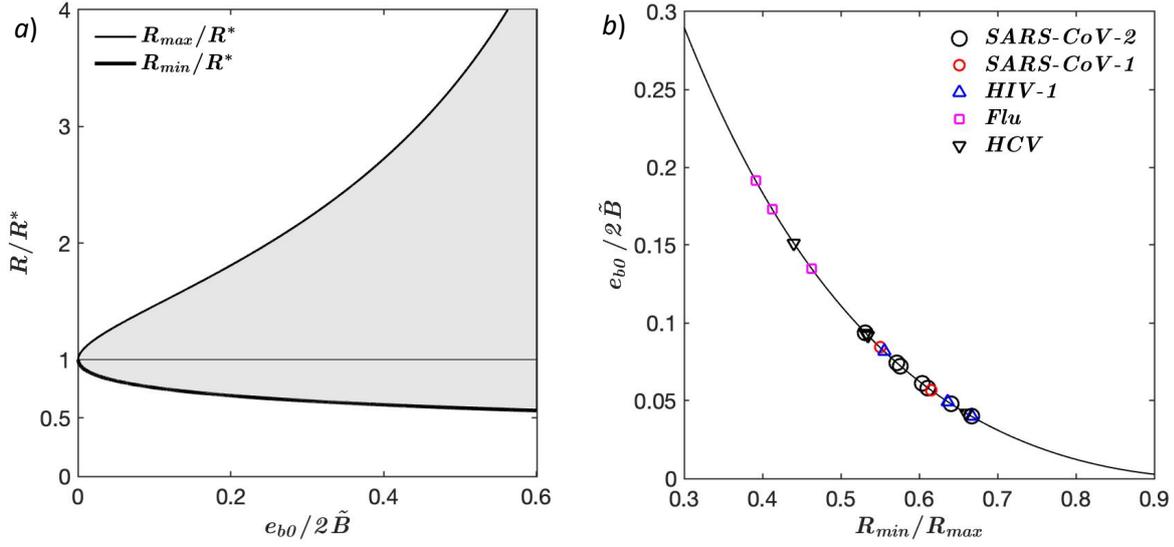

**Figure 3**: Bud size constraints as a function of input energy $e_{b0}$, from Eq. (8): *a)* Size constraints $R_{max}$ and $R_{min}$, relative to the spontaneous radius of curvature $R^*$, versus input energy ratio $e_{b0}/2\tilde{B}$; *b)* Predicted $e_{b0}/2\tilde{B}$ as a function of the ratio $R_{max}/R_{min}$ from the reported maximum/minimum virion size for virus species: SARS-CoV-1 & 2 [10-12], HIV-1 [13], Flu [14-15], and HCV (Hepatitis C) [16-17].

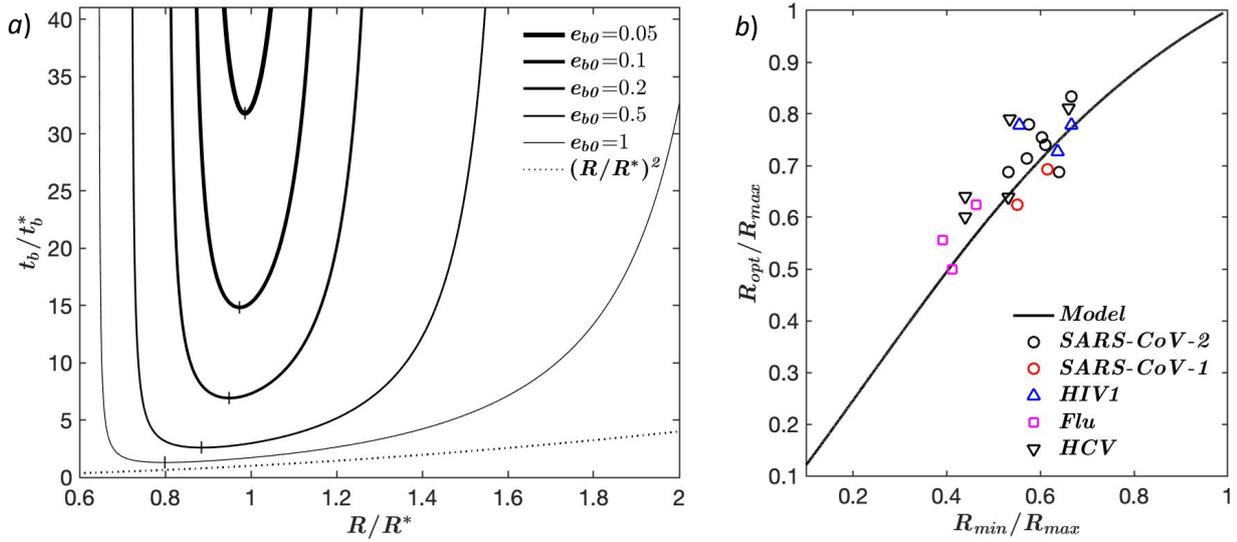

**Figure 4**: *a)* Dimensionless budding time $t_b/t_b^*$ versus dimensionless bud radius $R/R^*$ for various input energy $e_{b0}$, from Eq. (5), and (A8) and (A9) in *Appendix A*; The dashed line indicates $t_b/t_b^* = (R/R^*)^2$, valid for $e_{b0} \geq 5$; *b)* Predicted optimal-to-maximum radii $R_{opt}/R_{max}$ versus minimum-to-maximum radii $R_{min}/R_{max}$. The input energy is extracted from Figure 3b, and the predictions are compared against the observed median virion sizes for SARS-CoV-1 & 2 [10-12], HIV-1 [13], Flu [14-15], and HCV (Hepatitis C) [16-17].

Figure 4a plots the dimensionless budding time $t_b/t_b^*$, with $t_b^*$ the characteristic budding time, as a function of the dimensionless radius $R/R^*$ at various input energy $e_{b0}$. Eq. (5) reports the budding time $t_b$ as function of the speed factor $\alpha$, where the latter is then calculated numerically as function of the budding energy $e_b$ from Eq. (A8) and (A9) in *Appendix A*. The characteristic budding time is

$$t_b^* = \frac{R^{*2}}{D\alpha_\infty^2} \tag{11}$$

with $\alpha_\infty$ the maximum speed factor obtained for $e_b \to \infty$ and reported in Table A1 in *Appendix A.* The plots in Figure 4a are obtained for $\tilde{\rho} = 0.01$, where $\tilde{\rho} = 0.1$ gives nearly-identical plots, hence omitted. Note that, albeit $t_b/t_b^*$ appears not to be directly affected by $\tilde{\rho}$ within the explored range, $\tilde{\rho}$ affects the input energy $e_{b0}$, via Eq. (7), and $t_b^*$, via Eq. (11), where $\alpha_\infty$ depends on $\tilde{\rho}$ through Table

A1 (*Appendix A*). Each plot shows a minimum and maximum radius, calculated from Eq. (8), giving $t_b \to \infty$, as well as an optimal radius $R_{opt}$, for which $t_b = t_{b,min}$, the minimum budding time (vertical bar symbol). Larger $e_{b0}$ provides smaller $t_b$, intuitively, following the simple scaling law $t_{b,min}/t_b^* \approx 1.5/e_{b0}$ within the observed range. For $e_{b0} \geq 5$, the curves approach the simple scaling law $t_b/t_b^* \simeq (R/R^*)^2$ (dashed line in Figure 4a) for $R > R_{min}$, where $R_{opt} \simeq R_{min}$. This case, however, involves much higher input energy than the values extracted in Figure 3b, rendering this simple scaling inapplicable to the reported experimental observations.

From the values in Table 1 and the median value $e_{b0}/2\widetilde{B} \approx 0.05$, we have $e_{b0} \approx 0.16$ giving $t_{b,opt} \approx 10\, t_b^*$ from Figure 4a. From Tables 1 and A1 we then have $t_b^* \approx 0.15 - 2.93\ min$, for $\tilde{\rho} = 0.01 - 0.1$, respectively, from which $t_{b,opt} \approx 1.5 - 29.3\ min$. This range includes the experimentally observed $10 - 20\ min$ budding time [8].

Figure 4b reports the predicted ratio $R_{opt}/R_{max}$, extracted from Figure 4a, as a function of $R_{min}/R_{max}$, where $e_{b0}$ is extracted from Figure 3b. This figure compares the theoretical prediction with the observed optimal virion radius (from median values) for the species reported in Figure 3b. The close agreement between theory and experiments validates the proposed model under the assumption that budding provides the tightest constraint to virus size. This hypothesis is discussed in the next sub-sections.

The model considers the ideal condition of a fully formed capsid at the onset of budding. In some cases, the capsid is still forming when the first CS-TP binding occurs. This is particularly the case in retroviruses like HIV, where the capsid develops from the agglomeration of gags. In these cases, the attractive/repulsive forces between gags, and/or other capsid components, might affect $e_{b0}$. Thus, additional energetic terms may be needed in Eq. (7c) to account for it. Repulsive forces will reduce the effective $e_{b0}$, while attractive forces will increase it. Additionally, the diffusion of capsid components within the cytosol to reach the bud might also affect the kinetics of the process. The proposed model assumes that the process of capsid assembly is much faster than TP diffusion. This idealization could explain the larger deviation in Figure 4b for HIV-1.

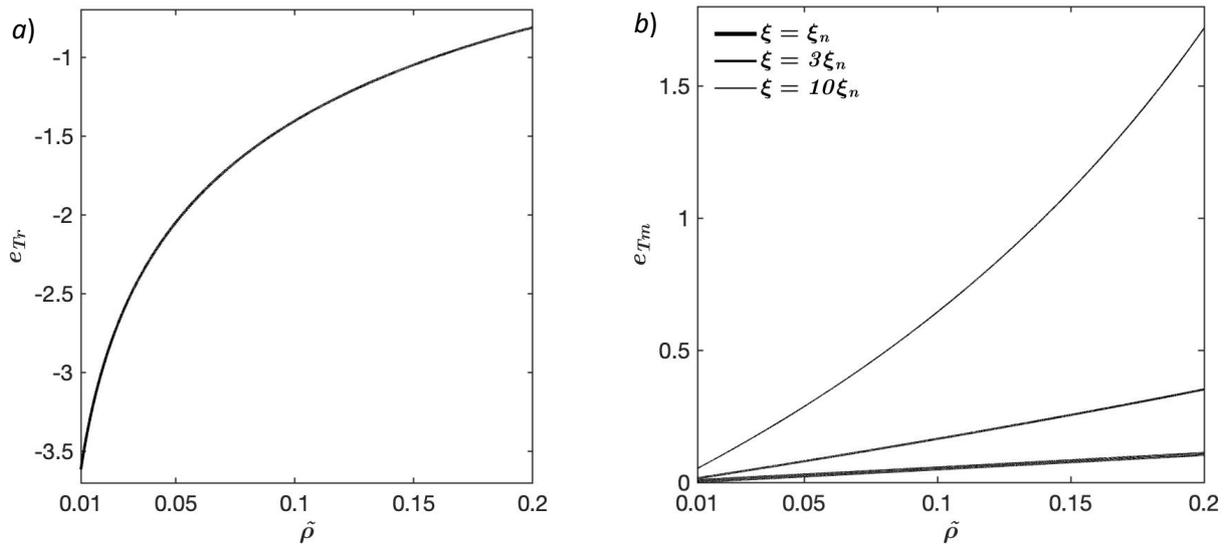

**Figure 5**: *a*) TP relocation energy $e_{Tr}$ (change in configurational energy) versus $\tilde{\rho} = \rho_0/\rho_b$ from Eq. (7d); *b*) membrane bending energy released by a TP joining the bud as a function of $\tilde{\rho}$ and $\xi$ from Eq. (7d), with $\xi_n = 0.2$.

Figure 5a reports the energy of TP relocation $e_{Tr}$, from Eq. (7d), as a function of $\tilde{\rho}$, for $\tilde{\rho} = 0.01 - 0.2$. This energy is negative as it constitutes a cost, with smaller $\tilde{\rho}$ requiring higher relocation costs due to the higher density gap between the bud and the resting membrane. This figure shows a logarithmic correlation between $\tilde{\rho}$ and $e_{Tr}$, within the observed range, evidencing the predominance of the third term on the right-hand side of Eq. (7d) over the others. Figure 5b reports the membrane bending energy released by a TP joining the bud, $e_{Tm}$, from Eq. (7e), as a function of $\tilde{\rho}$ for the same value range, and for $\xi = \xi_n$, $3\xi_n$, and $10\xi_n$, with $\xi_n = 0.2$ a nominal parameter value based on Eq. (7f) and Table 1. $e_{Tm}$ is positive since the release of this energy promotes budding. Figure 5b shows that $e_{Tm}$ is proportional to $\tilde{\rho}$ (TPs availability) and $\xi$, with the sensitivity of $e_{Tm}$ with respect to $\tilde{\rho}$ significantly amplified by $\xi$. From Eq. (7f) one can deduce that large $d_{TP}$ or low $\rho_b$ (large and sparsely distributed TPs) yield higher $e_{Tm}$. For the median values of the parameters adopted in Table 1, we can see in Figure 5b that the contribution of $e_{Tm}$ is relatively small ($< 0.1$) compared to $e_{Tr}$. Take now, for simplicity, $\tilde{g}_m = 0$, $\varepsilon_S = 0$, $\tilde{\rho} = 0.1$ (0.01), $e_{b0} = 0.16$, and $\xi = 0.2$. We have then $e_{Tr} = -1.4$ ($-3.62$), and $e_{Tm} = 0.053$ (0.0052), so that we can finally estimate the TP-capsid binding energy (in $kT$ units) as $e_{T_a} = 1.51$ (3.77).

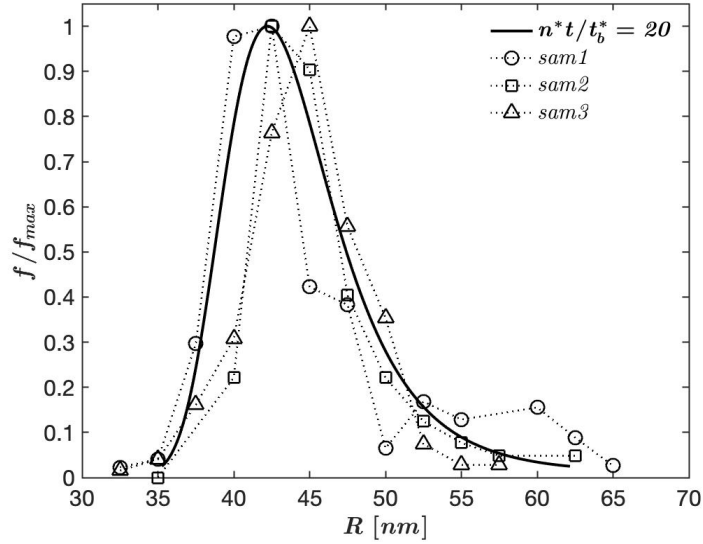

**Figure 6**: Distribution of the normalized statistical frequency $f/f_{max}$ versus virion radius $R$. The theoretical prediction (solid line) is generated with the proposed model, assuming hereditary size and exponential replication from Eq. (12) and (13). Taking $n^* t/t_b^* = 20$ as a fitting parameter, the prediction is compared against experimental observation on virion size polydispersity for SARS-CoV-2 [11], on three virus populations samples (sam1, 2, and 3).

*Exponential replication and virion size polydispersity*

Let us assume that virion size is hereditary, *i.e.* if a virion of radius $R$ infects the host, this will reproduce $n^*$ copies of itself having a radius that is very close to $R$.

Let us introduce the function $n(t, R)$ giving the number of virions having radius $R$ in the population at time $t$. Assuming that budding provides the tightest size constraint, the reproduction rate of virions having radius $R$, at the time $t$, becomes

$$\frac{\partial n(t,R)}{\partial t} = n^* \frac{n(t,R)}{t_b(R)} \qquad (12a)$$

By integrating Eq. (12a) with time, we have

$$n(t, R) = n_0(R) \exp\left(\frac{n^* t}{t_b(R)}\right) \qquad (12b)$$

where $n_0(R) = n(0, R)$. Consider now $n_0(R) = n_0$, $\forall R$, *i.e.* before any virus reproduction occurs the population has equal number of virions for any size. Now the total number of virions at the time $t$ is

$N(t) = \int_{R_{min}}^{R_{max}} n(t,R) dR$, via numerical integration, and the statistical frequency of virions having radius $R$ is $f(t,R) = n(t,R)/N(t)$, giving finally

$$f(t,R) = \frac{n_0}{N(t)} \exp\left(\frac{n^* t}{t_b(R)}\right) \tag{13}$$

From Eq. (13) we can now compare the statistical distribution of virion size with experimental observations. Figure 6 reports this comparison by taking three virus population samples (sam1, 2 and 3) from [11], where the vertical axis reports the normalized frequency with respect to its maximum in the population, $f/f_{max}$. Here, $f = f_{max}$ if $R = R_{opt}$. In the proposed model, one has to define the time of observation $t$, which here is taken as $t = 20\, t_b^*/n^*$. The parameter values used in this figure are $e_{b0} = 0.2$ and $R^* = 45\, nm$, estimated from the ratio $R_{min}/R_{max}$ of the samples [11], following the same procedure as in Figure 3b.

In the model here described, the virus population evolves toward the ideal condition at which all virions have radius $R_{opt}$, at $t \to \infty$, hence no polydispersity and $f(\infty, R) = \delta(R - R_{opt})$, with $\delta$ the Dirac delta. This condition is never reached in real virus populations, and this is likely due to the imperfect hereditariness of size, by which a virion might replicate virions having slightly different radii. A more comprehensive model should consider this deviation, but this is beyond the present scope.

*Size constraints of virus infection*

The experimental comparisons presented in the previous sections rely on the hypothesis that budding provides the tightest constraint to virion size. The other most important size-limiting process, in the life of a virion, is infection. This section discusses the size limitations introduced by infection, a process occurring via two main mechanisms, (i) receptor- and clathrin-mediated endocytosis, and (ii) membrane fusion. (i) requires the membrane of the host cell to wrap around the envelope to produce an endosome (Figure 7a), which later fuses with the enveloped membrane inside the cytosol to release RNA [27,3]. (ii) involves the fusion of the envelope membrane with that of the host cell prior to the creation of the endosome (Figure 7b) [27].

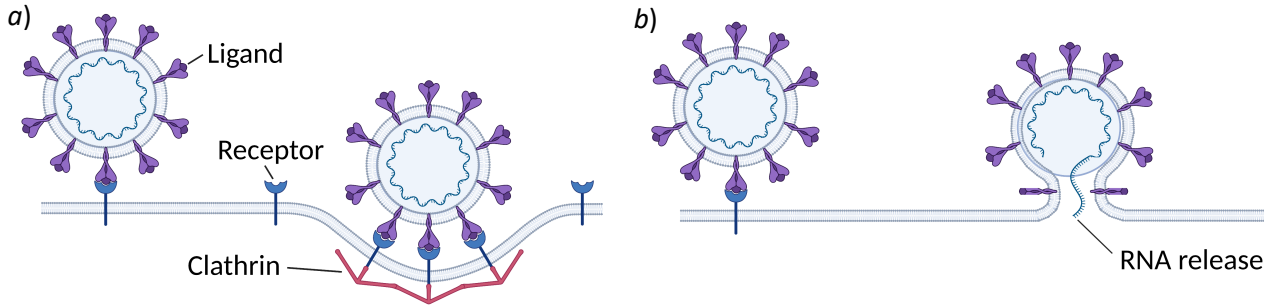

**Figure 7**: Infection mechanisms of enveloped viruses: *a*) Receptor- and clathrin-mediated endocytosis; *b*) Membrane fusion.

<u>Endocytosis</u>: The kinetics of the process is here controlled by the diffusion of receptors [3], and the proposed model applies by considering that now TPs stand for receptors (instead of virus ligands). In this case, Eq. (6) rewrites to

$$e_e \geq F_{\tilde{\rho}}(\alpha) \tag{14a}$$

where

$$e_e = e_{e0} - 2\tilde{B}\left(\frac{R^*}{R}\right)^2 - 2\tilde{B}_c\left(\frac{R^*}{R} - \frac{R^*}{R_c}\right)^2 \tag{14b}$$

is the *endocytosis energy* (the driving force for wrapping), with

$$e_{e0} = e_{RL} + \tilde{\rho}_c e_c - \tilde{\sigma}\varepsilon_S + 1 - \tilde{\rho} + \ln \tilde{\rho} \quad (14c)$$

the *input endocytosis energy*, and $F_{\tilde{\rho}}(\alpha)$ the kinetic function described in Eq. (6b-c). As previously discussed, TP diffusion is size independent, thus the value of $D$ is here again given in Table 1. In Eq. (14), $\tilde{B}_c$ is the bending rigidity of the clathrin coat (taken to be zero in the absence of clathrin), $R_c$ is its spontaneous radius of curvature, $e_{RL}$ is the receptor-ligand binding energy, $\tilde{\rho} = \rho_{r0}/\rho_b$, with $\rho_{r0}$ the surface density of receptors in the resting membrane, $\rho_b$ the density of receptors in the wrap (equivalent to the ligand density [3]), $e_c$ is the binding energy of the clathrin pit with the backing of the receptor [18], and $\tilde{\rho}_c = \rho_c/\rho_b$, with $\rho_c$ the surface density of clathrin pits.

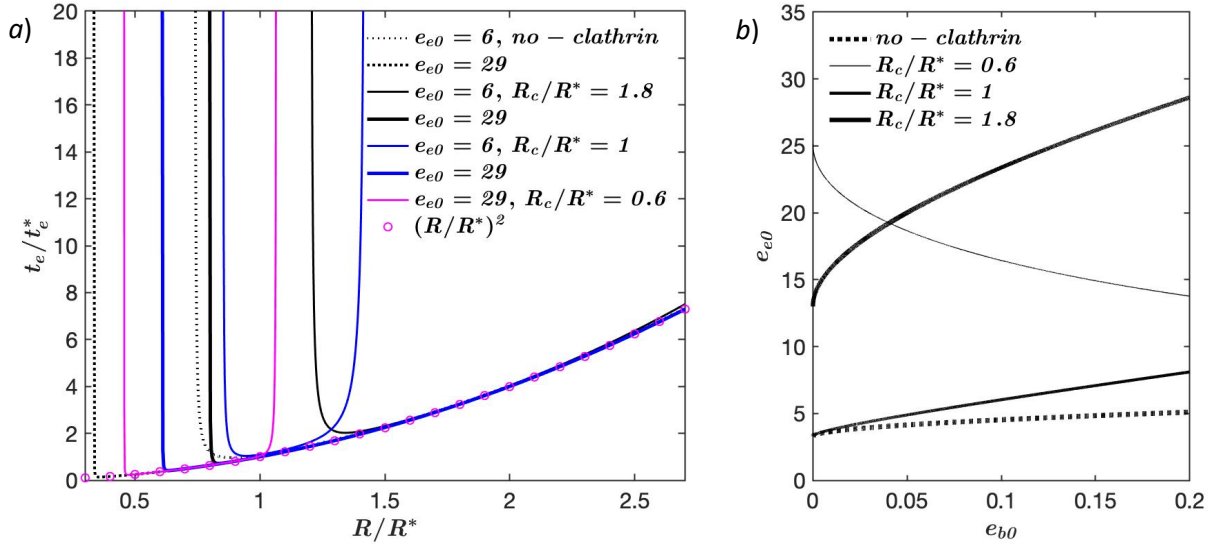

**Figure 8**: *a*) Dimensionless endocytosis (wrapping) time $t_e/t_e^*$ ($t_e^* = t_b^*$) versus virion size $R/R^*$, at various input endocytosis energy $e_{e0}$ and normalized spontaneous radius of curvature $R_c/R^*$. The results are obtained in the same way as in Figure 4a, with $e_b$ substituted with $e_e$ from Eq. (14b). The clathrin bending modulus is $B_c = 15 B$ (Table 1); *b*) Minimum $e_{e0}$ versus budding input energy $e_{b0}$, for which the highest selective pressure is provided by budding, with clarhtin (solid line), at various $R_c/R^*$, and with no clathrin (dashed line), from Eq. (16).

Figure 8a provides the dimensionless endocytosis (wrapping) time $t_e/t_e^*$, with $t_e^* = t_b^*$, versus dimensionless virion radius $R/R^*$, at various input energy $e_{e0}$, and normalized spontaneous curvature of clathrin $R_c/R^*$. The results are obtained in the same way as in Figure 4a, with $e_b$ substituted with $e_e$, from Eq. (14b). The dashed black lines represent the plots in the absence of clathrin ($\tilde{B}_c = 0$ and $\tilde{\rho}_c = 0$), while the solid black, blue, and magenta lines represent the plots in the presence of clathrin. For all these cases, the parameter values are taken from Table 1, with $\varepsilon_S = 0$ and the median values for $e_c$. In the absence of clathrin (receptor-mediated endocytosis), this gives $e_{e0} = 6.4 - 23.6$ and a median of $e_{e0} = 17.5$ (taking the median for $e_{RL}$ and averaged over $\tilde{\rho} = 0.01 - 0.1$). In the presence of clathrin, we have $e_{e0} = 12.13 - 29.3$ with a median of $e_{e0} = 23.24$. This is then rounded to $e_{e0} = 6 - 29$ in the figure. For $R_c/R^* = 0.6$, $e_{e0} = 6$ gives negative $e_e$, thus Eq. (14a) is violated and endocytosis cannot occur. For this reason, this plot is omitted. All the reported plots can be approximated by $t_e/t_e^* \approx (R/R^*)^2$ (magenta circles) for $R_{min} < R < R_{max}$. Also, the plots consider $\tilde{\rho} = 0.01$, where $\tilde{\rho} = 0.1$ gives nearly identical plots, hence omitted. As observed by [3] and confirmed in this figure, endocytosis provides size limitations, where virions having a radius smaller than $R_{min}$ are denied entry due to the excessive (bending) energy barrier. The same authors also predicted a maximum radius $R_{max}$ due to receptor depletion. However, as calculated by [18], and shown in this figure, the presence of clathrin also creates an $R_{max}$, and generally provides a tighter constraint to the minimum and maximum virion radii. The input energy $e_{e0}$ appears here not to affect the endocytosis time significantly, so long that the virion size $R$ is within its limits. However, $e_{e0}$ has a strong influence on virion size limitations, particularly in the presence of clathrin, and more so for $R_c \leq R^*$.

Taking $R \approx 50\ nm$, we have $R/R^* \approx 1$, from which $t_e \approx t_b^* \approx 0.15 - 2.93\ min$. This is in qualitative agreement with the $2 - 58\ s$ optimal wrapping time calculated by [3].

Let us now analyze the condition for which budding provides the tightest selective pressure to the size of the virions. Let us assume that we are at the limit for budding spontaneity, i.e. $\alpha = 0$ and $t_b \to \infty$, giving $e_b = 0$. From Eq. (7a), this condition gives that $R = R_{min}$ or $R = R_{max}$ from Eq. (8b-c), i.e.

$$\frac{R^*}{R} = 1 \pm \sqrt{\frac{e_{b0}}{2\tilde{B}}} \tag{15}$$

Substitution of Eq. (15) into (14b), with condition (14a), gives

$$\frac{e_{e0}}{2\tilde{B}} \geq \left(1 \pm \sqrt{\frac{e_{b0}}{2\tilde{B}}}\right)^2 + \frac{B_c}{B}\left(1 \pm \sqrt{\frac{e_{b0}}{2\tilde{B}}} - \frac{R^*}{R_c}\right)^2 \tag{16}$$

where the tightest limitation comes with $\pm$ substituted with $+$.

Figure 8b reports the minimum input energy $e_{e0}$ satisfying Eq. (16) as a function of $e_{b0}$ and $R_c/R^*$, and with $B_c = 0$ and $B_c = 15\ B$ from Table 1. The dashed lines report the case of no-clathrin. In the range of the parameter values reported in Table 1, the range of $e_{e0}$ discussed in Figure 8a, and in the absence of clathrin, Eq. (16) is always satisfied. In the presence of clathrin, Eq. (16) is satisfied for all cases apart from that of $R_c/R^* = 0.6$ and $1.8$ for certain values of $e_{b0}$. For the adopted median parameter values, $R_c/R^* = 1$, and Eq. (16) is always satisfied. Thus, we can conclude that, for the most representative cases of infection via endocytosis, the tightest size constraint is dictated by budding.

*Membrane fusion*: This process involves virus unwrapping, as shown in Figure 7b and Figure A1c, where the membrane of the virion fuses with that of the host and transits from a spherical envelope to a flat configuration. In this case, the ligands of the virion, one by one, detach from the nucleocapsid and then diffuse away toward remote regions in the host membrane. It starts with the formation of a fusion pore, via molecular reconfiguration of receptor-ligand bonds (proteolytic cleavage). In this case, TP stands for the ligands of the virus. The propagation of the fusion pore is described by the solution to the transient problem reported in *Appendix B*, under the assumption of a diffusion-limited regime, where the process continues thanks to the detachment and diffusion of TPs. This is described by

$$e_f \geq F_{\tilde{\rho}_f}(\alpha) \tag{17a}$$

with

$$e_f = e_{f0} + 2\tilde{B}\left(1 - \frac{R^*}{R}\right)^2 \tag{17b}$$

the fusion energy,

$$e_{f0} = e_{RL}\gamma - \tilde{g}_m - e_T + \tilde{\sigma}\varepsilon_S - (1 + \tilde{\rho}_f + \ln \tilde{\rho}_f) \tag{17c}$$

the *input fusion energy*, and

$$F(\alpha)_{\tilde{\rho}} = (\tilde{\rho}_f + 1)f(\alpha) + \ln\left[1 + \left(1 + \frac{1}{\tilde{\rho}_f}\right)f(\alpha)\right] \tag{17d}$$

the kinetic function of fusion, with $\tilde{\rho}_f = \rho_{f0}/\rho_b$, where $\rho_{f0}$ is the equilibrium TP density in the host membrane, and $f(\alpha)$ is given by Eq. (6c). In Eq. (17), $\gamma = S_f/S_b$ with $S_f$ the area of the nucleated fusion pore and $S_b = 4\pi R^2$ the surface of the envelope. The speed factor $\alpha$ identifies the velocity of the fusion process. As specified in *Appendix B*, for $e_f \to \infty$ we have that $\alpha \to \infty$, *i.e.* TP diffusion can be indefinitely fast (unlike budding and endocytosis). Due to the unlimited speed factor, TP diffusion can become faster than membrane relaxation or equally fast. In this case, the fusion time $t_f$ calculated in *Appendix B* is inaccurate since the proposed model relies on the assumption that viscoelastic relaxation is much faster than TP diffusion. This model neglects again the bending energy of the membrane outside the envelope and, thus, also the line energy of the rim. This approximation is again based on the hypothesis that the membrane outside the wrap assumes a catenoid-like configuration [4].

Figure 9a plots the dimensionless fusion time $t_f/t_f^*$, with $t_f^*$ the characteristic fusion time, versus $R/R^*$, at various $e_{f0}$ and $\tilde{\rho}_f$. The fusion time is given by Eq. (5), with $t_b$ substituted with $t_f$, as function of the speed factor $\alpha$. The latter is then calculated numerically as function of the fusion energy $e_f$ from Eq. (17) and (B7) in *Appendix B*. The characteristic fusion time is

$$t_f^* = \frac{R^{*2}}{D} \tag{18}$$

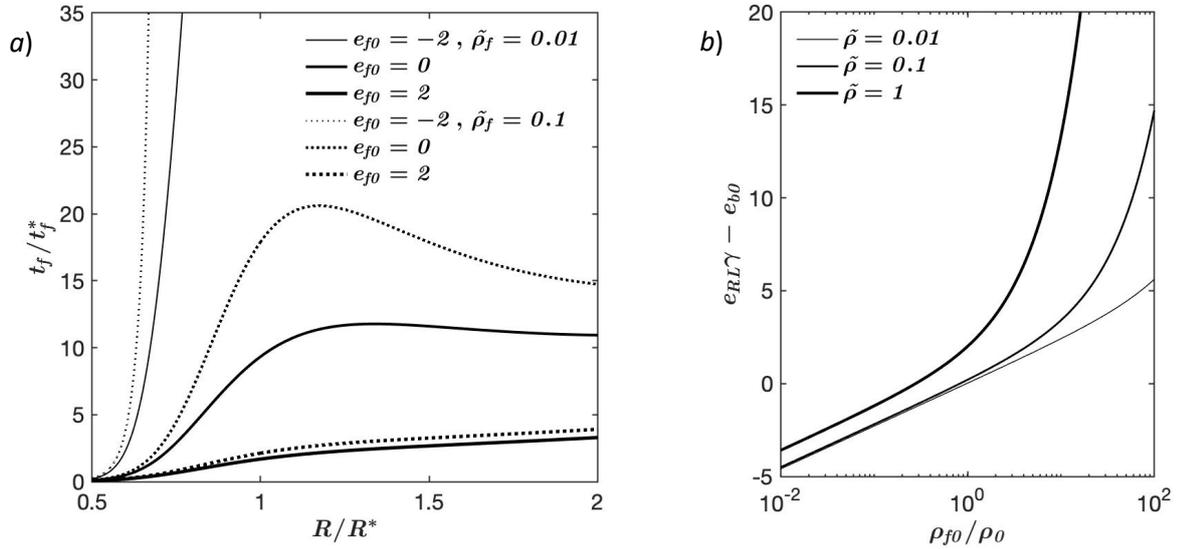

**Figure 9**: *a)* Dimensionless fusion (unwrapping) time $t_f/t_f^*$ versus virion size $R/R^*$ at various $e_{f0}$ and $\tilde{\rho}_f$, from Eq. (5) (with $t_b$ substituted with $t_f$) and (17), and from Eq. (B7) from *Appendix B*; *b)* Minimum fusion pore energy $e_{RL\gamma}$ required to promote membrane fusion at the limiting conditions of budding, as described in Eq. (19), for various equilibrium TP densities $\tilde{\rho}$ in the mother cell

As shown in this figure, $t_f$ is smallest for $R \approx R_{min}$ or $R \approx R_{max}$, and largest for $R$ near $R^*$, showing an opposite correlation compared to $t_b$ and $t_e$. This is intuitive since the bending energy is now a promoter of fusion, rather than an energy barrier. Also, the fusion time and size-dependent kinetics appear here to be highly sensitive to the input energy $e_{f0}$.

It is important to notice that virus infection only requires the injection of RNA, which can occur as soon as the pore is large enough, compared to the envelope size. Albeit completed unwrapping is far from necessary for infection, we can still assume that the infection time is proportional to $t_f$. It is interesting to notice that fusion can occur with negative $e_f$ if $R \ll R^*$. From the parameter values in Table 1 and Eq. (18), we can obtain $t_f^* = 0.25\ s$. The relaxation time of the cell membrane is $t_r = 0.5 - 0.75\ s$ (measured on a red blood cell [28]), thus the prediction reported in Figure 9a is reliable only for $e_f \leq 0$ and $R/R^* \geq 0.7$. *I.e.* for the reported smallest values of $t_f/t_f^*$ (for $R \to 0.5$), TP diffusion is faster than membrane relaxation, hence, the latter controls the kinetics of the process and the proposed model is inapplicable.

From Figure 9a we can deduce that $t_f/t_f^* \approx 0 - 100$, for the observed parameter values, thus $t_f \approx 1 - 10\ s$.

Let us now analyze the hypothesis of budding as the tightest size constraint. Because the statistical size frequency of virions has the median at an intermediate radius, and lower virion count for larger and smaller radii, it is intuitive to consider that membrane fusion has limited influence on the size polydispersity of a virus population by looking at Figure 9a. However, it is useful to analyze the critical conditions required for the spontaneity of membrane fusion. By equating Eq. (7) with (17), this condition becomes

$$e_{RL}\gamma - e_b \geq \tilde{\rho}\left(1 + \frac{\tilde{\rho}_f}{\tilde{\rho}}\right) + \ln\left(\frac{\tilde{\rho}_f}{\tilde{\rho}}\right) \tag{19}$$

Because $e_b \leq e_{b0}$, from Eq. (7), we can deduce that $R = R^*$ produces the worst-case scenario in Eq. (19) so that one can simply substitute $e_b$ with $e_{b0}$ in this equation.

Figure 9b reports the minimum $e_{RL}\gamma - e_{b0}$ required to favor fusion, from the right-hand side of Eq. (19) and for various values of $\tilde{\rho}$ and $\tilde{\rho}_f/\tilde{\rho} = \rho_{f0}/\rho_0$. Noteworthy, $\rho_0$ is the equilibrium density of ligands in the mother cell, where the virion is generated, while $\rho_{f0}$ is that in the infected cell. We can observe that a larger $\rho_{f0}/\rho_0$ raises the bar for spontaneous fusion, while a smaller $\rho_{f0}/\rho_0$ can facilitate fusion. *I.e.* according to this model, the first infection of a cell is far more likely to succeed and proceed at a high rate than the infection of a cell that has already been infected multiple times. Considering now $\rho_{f0} = \rho_0$, Eq. (19) reduces to $e_{RL}\gamma - e_{b0} \geq 2\tilde{\rho}$. For the range of parameter values adopted in Table 1, for $e_{b0}$ and $e_{RL}$, one can estimate $\gamma \geq 0.0048 - 0.086$, with median values giving $\gamma \geq 0.014$. *I.e.*, the spontaneous formation and propagation of the fusion pore can require up to 8.6% of the envelope surface to be covered with receptor-ligand bonds. Take now the fusion pore surface $S_f = \pi r_f^2$, with $r_f$ its radius, and $S_f = \gamma\ S_b = \gamma\ 4\pi R^2$, with $R \approx 50\ nm$ from Table 1. We can then estimate the pore radius as $r_f = 2R\sqrt{\gamma} = 6.93 - 29.33\ nm$, with median value at $r_f = 11.62\ nm$. By comparing this with the ligand spacing $l_b = 14\ nm$ in Table 1, we can deduce that the fusion pore requires commonly only one receptor-ligand bond to nucleate, and in some cases, it can require up to 3-4 bonds.

## 4. Conclusions

The proposed simple model provides an energetic analysis to derive size constraints and size-dependent budding time for an enveloped virus. This compares well with experimental observations on the optimal size (statistical median), giving fastest budding, in virus populations for SARS-CoV-1 & 2 (coronavirus), HIV-1, Flu, and HCV (hepatitis C). The model also shows a good prediction of the size polydispersity for a virus population, via a simple exponential replication model based on perfect size-hereditariness, for SARS-CoV-2. The same comparison can be simply extended to the other species of enveloped virus analyzed. These comparisons rely on the assumption that budding provides the tightest size constraint. This hypothesis is verified by analyzing the size and energy limitations introduced by infection via receptor- and clathrin-mediated endocytosis, and via membrane fusion. Furthermore, as discussed in this manuscript, the timescale for virus replication (via budding) is $\sim 10\ min$, while that of virus infection goes from $\sim 10\ s$, for membrane fusion, to $\sim 1\ min$, for endocytosis. From this, one can observe that the life cycle of a virus can be much more impacted by the kinetics of replication than that of infection. This observation could shed light on the biophysical mechanisms involved in virus infectivity. The experimental validation presented in this paper also suggests that, for the analyzed virus species, budding is mediated by TP diffusion. The model is constructed to organize its parameters in a hierarchical fashion, so that the dimensionless budding time $t_b/t_b^*$ only depends on a small number of dimensionless parameters, namely, the virion size $R/R^*$, the availability of TPs $\tilde{\rho}$, and the input energy $e_{b0}$. While $R/R^*$ varies across a small range, $\tilde{\rho}$ and $e_{b0}$ are observed to span across at least one order of magnitude. This paper also provides a simple scaling between the budding time and these quantities. The proposed model provides a valuable tool to correlate molecular-scale properties (*e.g.* TP-capsid binding energy) of a single enveloped virion with the size polydispersity of a virus population. Because molecular-scale properties are often difficult to measure (*e.g.* to the author's knowledge, no experimental data is available for TP-capsid adhesion), this model can be used in reverse to extrapolate these properties from statistical observations on size polydispersity. Finally, the proposed model can inspire biomolecular strategies to limit virus replication by reducing the budding energy, thereby increasing the budding time of a virus population.

## Acknowledgments

This work was supported by the New Frontiers in Research Funds – Exploration (NFRFE-2018-00730), and the Natural Sciences and Engineering Research Council of Canada (NSERC) (RGPIN-2017-04464, and ALLRP554607-20).

## Appendix A: Diffusion-mediated budding (and endocytosis)

The proposed model assumes that TP diffusion controls the kinetics of the process, as discussed by [8]. In axial symmetry, the radial diffusion $j$ is described by Fick's second law as

$$j = -D\rho_{,r} \tag{A1a}$$

or

$$j = -\rho D\mu_{,r} \tag{A1b}$$

with $r$ the distance from the axis of symmetry, and $D$, $\rho$ and $\mu = 1 + \ln(\rho/\rho_0)$ the diffusivity, surface density and (dimensionless) chemical potential of TPs in the membrane, respectively. In this case, TP conservation imposes

$$\dot{\rho} = -\frac{1}{r}(rj)_{,r} \tag{A2}$$

everywhere in the membrane, where $\dot{\rho} = \partial\rho/\partial t$ is the local rate of change of TP concentration, with $t$ time. Substitution of (A1) into (A2) provides a partial differential equation in the function $\rho(t,r)$, subjected to the boundary conditions $\rho(t,\infty) = \rho_0$ and initial conditions $\rho(0,r) = \rho_0$, with $\rho_0$ the equilibrium TP density in the unperturbed membrane (prior to budding or far away from the bud). The solution to this problem is given by [3]

$$\rho = \rho_0 + C\, E\left(\frac{r^2}{4Dt}\right) \tag{A3}$$

with $C$ an integration constant and $E(x) = \int_x^\infty \exp(-u)\, du/u$ the exponential integral.

The total number of TPs in the system is

$$n_{TP} = \rho_b S_b + 2\pi \int_{r_1}^\infty \rho\, r dr \tag{A4}$$

with $\rho_b$ the (constant) TP density on the bud, $S_b$ the surface of the bud, and $r_1$ the distance between the bud rim and the axis of symmetry (Figure 1a). The assumption of constant $\rho_b$ (and spacing, $l_b = \rho_b^{-1/2}$, [18]) derives from observations on ligand density and spacing for Influenza-A virus [29] and

SARS-CoV-2 [30]. In this case, TP conservation imposes $\dot{n}_{TP} = 0$ on Eq. (A4), which, considering $\dot{S}_b = 2\pi r_1 \dot{r}_1$ and (A1-A3), gives

$$(\rho_b - \rho_1)\dot{r}_1 + j_1 = 0 \tag{A5}$$

where $\rho_1$ and $j_1$ are the TP density and flux at the bud rim $r = r_1$, respectively. Because the condition at Eq. (A5) must be satisfied at any instant, the substitution of (A3) into (A5) gives the kinetic law for the moving boundary at Eq. (4).

By substituting (A1), (A3) and (4) into (A5), the integration constant computes to

$$C = \frac{\alpha^2(\rho_b - \rho_0)}{\alpha^2 E(\alpha^2) - \exp(-\alpha^2)} \tag{A6}$$

To calculate $\alpha$, one has to define the thermodynamics of the problem. The free energy of the system is described by Eq. (1), and its rate of change, from the substitution of (A1), (A2) and (A5) into it, gives

$$\frac{\dot{\psi}}{2\pi kT} = \left[ -\rho_b e_T - g_m + \sigma \varepsilon_S + \rho_1 - \rho_b + \rho_b \ln\left(\frac{\rho_b}{\rho_1}\right) + 2B\left(\frac{1}{R} - \frac{1}{R^*}\right)^2 \right] r_1 \dot{r}_1 - \frac{\dot{q}}{2\pi kT} \tag{A7a}$$

with

$$\frac{\dot{q}}{2\pi kT} = \int_{r_1}^{\infty} D\rho(\mu_{,r})^2 r\,dr \tag{A7b}$$

The term in Eq. (A7b) is the energy dissipation due to the diffusive transport of TP across the membrane, from remote locations to the bud rim. To ensure the spontaneity of the process, one needs to ensure a continuous free energy reduction, i.e. $\dot{\psi} \leq 0$ (second law of thermodynamics).

Figure A1a gives the distribution of $\rho$ across the membrane, from Eq. (A6) substituted into (A3) and with $\rho_b > \rho_0$. For budding and endocytosis (Figure A1b-c), $\rho_1 < \rho_0$ is required to prompt the flux of TP toward the bud. From Eq. (A7b) one can deduce that $\dot{q} \geq 0$ is always satisfied. At this point, budding spontaneity requires that the term in [ ], in Eq. (A7a), satisfies [ ] $\leq 0$. This condition, from the substitution of Eq. (A6) into (A3) and the result into (A7a), ultimately provides Eq. (6).

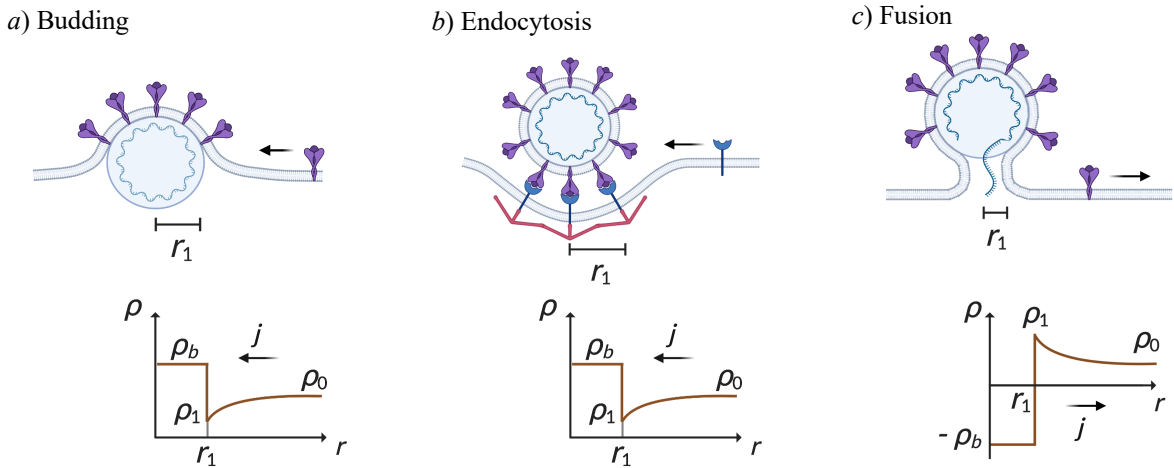

**Figure A1**: Schematic (top) of budding *a*), endocytosis *b*) and fusion *c*), with corresponding TP density distribution (bottom). The density distribution promotes the TP flux from remote regions toward the bud, for budding and endocytosis, and from the bud toward remote regions for fusion.

The kinetic constant is then obtained from $\alpha = F_{\tilde{\rho}}^{-1}(e_b)$, by inverting Eq. (6a). This is here done numerically, giving

$$\alpha = \alpha_\infty \tilde{\alpha}(e_b) \tag{A8}$$

with $\alpha_\infty = F_{\tilde{\rho}}^{-1}(\infty)$ the maximum speed factor, and $\tilde{\alpha}(e_b)$ the normalized speed factor giving $\tilde{\alpha}(0) = 0$ and $\tilde{\alpha}(\infty) = 1$. The latter is fitted to the function

$$\tilde{\alpha}(e_b) = 1 - \exp(-k_1 e_b - k_2\sqrt{e_b}) \tag{A9}$$

The values for $\alpha_\infty$ and the fitting constants $k_1$ and $k_2$, are given in Table A1 for $\tilde{\rho} = 0.01$ and $\tilde{\rho} = 0.1$. The fitting accuracy of Eq. (A9), with the coefficients in Table A1, is $R^2 = 1$ for both the adopted values of $\tilde{\rho}$.

| $\tilde{\rho}$ | $\alpha_\infty$ | $k_1$ | $k_2$ |
|---|---|---|---|
| 0.01 | 0.0415 | 0.7443 | 0.7002 |
| 0.1 | 0.1846 | 0.8043 | 0.6326 |

**Table A1**: Fitting parameters for Eq. (A9).

The budding time $t_b$, expressed in Eq. (5) as a function of $\alpha$, can be rewritten in dimensionless form as

$$\frac{t_b}{t_b^*} = \left(\frac{R}{R^*}\right)^2 \frac{1}{\tilde{\alpha}(e_b)^2} \tag{A10}$$

with $t_b^*$ the characteristic budding time taken from Eq. (11), and $\tilde{\alpha}(e_b)$ from Eq. (A9). The dimensionless budding time in Eq. (A10) is finally reported in Figure 4a.

## Appendix B: Diffusion-mediated membrane fusion

Assuming TP diffusion controls again the kinetics of the process, in axial symmetry, the radial diffusion $j$ is described by Fick's second law as in Eq. (A1) and (A2). The TP (ligand) density distribution $\rho_f(t,r)$ is again given by (A3) with $\rho$ and $\rho_0$ substituted with $\rho_f$ and $\rho_{f0}$.

The total number of TP now is

$$n_{TP} = -\rho_v S_v + 2\pi \int_{r_1}^{\infty} \rho_f\, r dr \tag{B1}$$

where $\rho_v$ and $S_v$ are the TP density and surface of the virion, and the minus on the first term on the right-hand side is due to the extraction of TPs from the envelope to then join the resting membrane (Figure A1). TP conservation imposes again $\dot{n}_{TP} = 0$ on Eq. (B1), which, considering $\dot{S}_b = 2\pi r_1 \dot{r}_1$ and (A1-A3), gives

$$-(\rho_v + \rho_1)\dot{r}_1 + j_1 = 0 \tag{B2}$$

where $\rho_1$, $r_1$ and $j_1$ are now associated with the rim of the fusion pore. The latter advances again according to the kinetic law given by Eq. (4)

By substituting (A1), (A3) and (4) into (B2), the integration constant computes to

$$C = \frac{\alpha^2(\rho_{f0}+\rho_v)}{\exp(-\alpha^2)-\alpha^2 E(\alpha^2)} \tag{B3}$$

To calculate $\alpha$, one has to again define the thermodynamics of the problem. The free energy of the system is now described by

$$\frac{\psi}{kT} = \left[g_m + \rho_v e_T - \sigma \varepsilon_S - \rho_v \ln\left(\frac{\rho_v}{\rho_0}\right) - 2B\left(\frac{1}{R} - \frac{1}{R^*}\right)^2\right] S_v + 2\pi \int_{r_1}^{\infty} \rho_f \ln\left(\frac{\rho_f}{\rho_{f0}}\right) r dr \tag{B4}$$

and its rate of change, from the substitution of (A1), (A2) and (B2) into it, gives

$$\frac{\dot{\psi}}{2\pi kT} = \left[ g_m + \rho_v e_T - \sigma \varepsilon_S - 2B\left(\frac{1}{R} - \frac{1}{R^*}\right)^2 + \rho_1 + \rho_v + \rho_v \ln\left(\frac{\rho_1}{\rho_v}\right) \right] r_1 \dot{r}_1 - \frac{\dot{q}}{2\pi kT} \tag{B5}$$

with $\dot{q}$ the energy dissipated by TP transport, given by Eq. (A7b). To ensure the spontaneity of the process, one needs to again ensure continuous free energy reduction, *i.e.* $\dot{\psi} \leq 0$ (second law of thermodynamics). Here, one can again ensure spontaneity by imposing $[\ ] \leq 0$ for the terms within $[\ ]$ in Eq. (B5). This then provides the condition in Eq. (17).

The fusion time $t_f$, similarly to the budding time, can be written in dimensionless form as

$$\frac{t_f}{t_f^*} = \left(\frac{R}{R^*}\right)^2 \frac{1}{\alpha(e_f)^2} \tag{B6}$$

and plotted in Figure 9b, with $t_f^*$ the characteristic fusion time given by Eq. (18). Here, the relation $\alpha(e_f)$ in Eq. (B6) is calculated from the numerical solution of Eq. (17d).